\documentclass[]{llncs}

\usepackage[utf8]{inputenc}
\usepackage[english]{babel}

\usepackage[dvipdfmx]{graphicx}
\usepackage{amsmath}
\usepackage{amssymb}
\usepackage{newtxtext,newtxmath}
\usepackage{bm}
\usepackage{listings}
\usepackage{color}
\usepackage{url}
\usepackage{afterpage}
\usepackage{subfigure}
\usepackage{wrapfig}
\usepackage{float}
\usepackage{multirow}
\usepackage{bussproofs}
\usepackage{intmacros}
\usepackage{algorithmic}

\usepackage{article}

\definecolor{clKeys}{rgb}{0,0,1}
\definecolor{clIdentifier}{rgb}{0,0,0}
\definecolor{clComments}{rgb}{0.3,0.3,0.3}
\definecolor{clString}{rgb}{0,0.5,0}
\definecolor{clLogic}{rgb}{0.5,0,0}
\begin{document}

\mainmatter

\title{HySIA: Tool for Simulating and Monitoring\\ Hybrid Automata Based on Interval Analysis}
\author{Daisuke Ishii\inst{1} \and Alexandre Goldsztejn\inst{2}}
\institute{University of Fukui, Japan. ~ \email{dsksh@acm.org} \and
  CNRS/LS2N, France. ~ \email{alexandre.goldsztejn@gmail.com}}

\maketitle

\begin{abstract} 
    We present HySIA: a reliable runtime verification tool for nonlinear hybrid automata (HA) and signal temporal logic (STL) properties.
    HySIA simulates an HA with interval analysis techniques so that a trajectory is enclosed sharply within a set of intervals. 
    Then, HySIA computes whether the simulated trajectory satisfies a given STL property; the computation is performed again with interval analysis to achieve reliability.
    Simulation and verification using HySIA are demonstrated through several example HA and STL formulas.
\end{abstract}


\section{Introduction}

Runtime verification of hybrid systems is realized with monitoring tools (e.g., \cite{Donze2010a,Annpureddy2011}) and statistical model checkers (e.g., \cite{David2012,Zuliani2013,Wang2015}) based on numerical simulation. These tools are practical because they can utilize de-facto standard environment (e.g., MATLAB/Simulink) for modeling and simulating industrial systems that are large and nonlinear.
However, their underlying numerical computation is unreliable due to numerical errors and can result in incorrect verification results. 
Conversely, computation of a rigorous overapproximation of a behavior (or a reachable region) suffers a trade-off in the precision of resulting enclosures and computational costs~\cite{SpaceEx2011,Chen2013,Kong2015}. A large \emph{wrapping effect} may occur when a model involves nonlinear expressions.

This paper presents the HySIA tool, a reliable simulator and verifier for hybrid systems.
HySIA supports nonlinear hybrid automata (HA) whose ODEs, guards, and reset functions are specified with nonlinear expressions. It assumes a deterministic class of HA; a transition to another location happens whenever a guard condition holds.
The main functionalities of HySIA are the following:

\subsubsection*{Simulation.}
HySIA simulates an HA based on interval analysis;
it computes an overapproximation of a bounded trajectory (or a set of trajectories) that is composed of \emph{boxes} (i.e., closed interval vectors) and \emph{parallelotopes} (linear transformed intervals) using our proposed method~\cite{GI2016}.
The computation can also be regarded as reachability analysis.
Intensive use of interval analysis techniques distinguishes HySIA from other reachability analysis tools.
First, the simulation process carefully reduces the {wrapping effect} that can expand an enclosure interval. As a result, HySIA is able to simulate an HA for a greater number of steps than other overapproximation-based tools; e.g., it can simulate a periodic bouncing ball for more than a thousand steps.
Second, HySIA relies on the soundness of interval computation so that the resulting overapproximation is verified to contain a theoretical trajectory.
This verification may fail, e.g., when an ODE is \emph{stiff} or when a trajectory and a guard are close to tangent, resulting in an enclosure too large to enable any inference.
Due to this \emph{quasi-complete} manner, the simulation process of HySIA performs efficiently whenever a numerically manageable model is given.

\subsubsection*{Monitoring.}
HySIA takes a temporal property as an input and monitors whether a simulated trajectory of an HA satisfies the property. Otherwise, HySIA is able to compute a \emph{robustness} signal~\cite{Donze2010} for the property.
The monitoring process~\cite{Maler2003,Donze2010} for \emph{signal temporal logic} (STL) formulas is extended to handle overapproximation of trajectories.
The soundness of interval computation is again utilized here to evaluate the logical negation against an overapproximated trajectory~\cite{Ishii2016IEICE}.

\subsection{Related Work}

Several tools for simulation and reachability analysis of hybrid systems based on interval analysis have been developed, including Acumen~\cite{Duracz-SNR-2016}, dReach~\cite{Kong2015}, Flow*~\cite{Chen2013}, iSAT-ODE~\cite{Eggers2012}, and HySon~\cite{Bouissou2012}.
They enclose a trajectory with a sequence of numerical processes e.g., for ODE integration, guard detection, and discrete jump computation; therefore, they suffer from the wrapping effect because each process outputs a result as an explicit overapproximation.
In contrast, HySIA regards a continuous and a discrete change as a composite function and evaluates it with a single overapproximation process.
Comparison results between HySIA and Flow* or dReach are reported in \cite{GI2016} or \cite{Ishii2015-NSV}, respectively.
With sufficiently small uncertainties, HySIA outperforms other tools.
Conversely, Flow* aims at handling models with large uncertainties by enclosing a state with a higher-order representation and outperforms HySIA in this respect.

SReach~\cite{Wang2015} and ProbReach~\cite{Shmarov2015} are statistical model checkers based on interval analysis. SReach cooperates with dReach to exploit its reachability analysis results.
ProbReach evaluates the continuous density function with interval analysis computation.
RobSim~\cite{Fainekos2009} utilizes an interval-based integration method for checking the correctness and robustness of the numerical simulator of Simulink.
HySIA can be involved within the frameworks proposed in these tools.

\section{Implementation}

\begin{wrapfigure}[10]{r}{0.5\textwidth}
  \centering
  \vspace{-2.5em}
  \includegraphics[width=\linewidth]{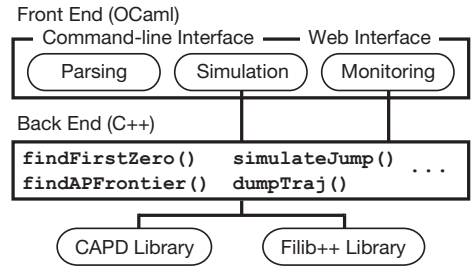}
  \vspace*{-2em}
  \caption{System structure of HySIA.}
  \label{f:structure}
\end{wrapfigure}

The brief structure of HySIA is shown in Figure~\ref{f:structure}.
The front end of HySIA is implemented in OCaml and the back end is implemented in C++.
The source repository is available at \url{https://github.com/dsksh/hysia}.

\begin{figure}[t]
  \centering
    \includegraphics[width=.65\linewidth]{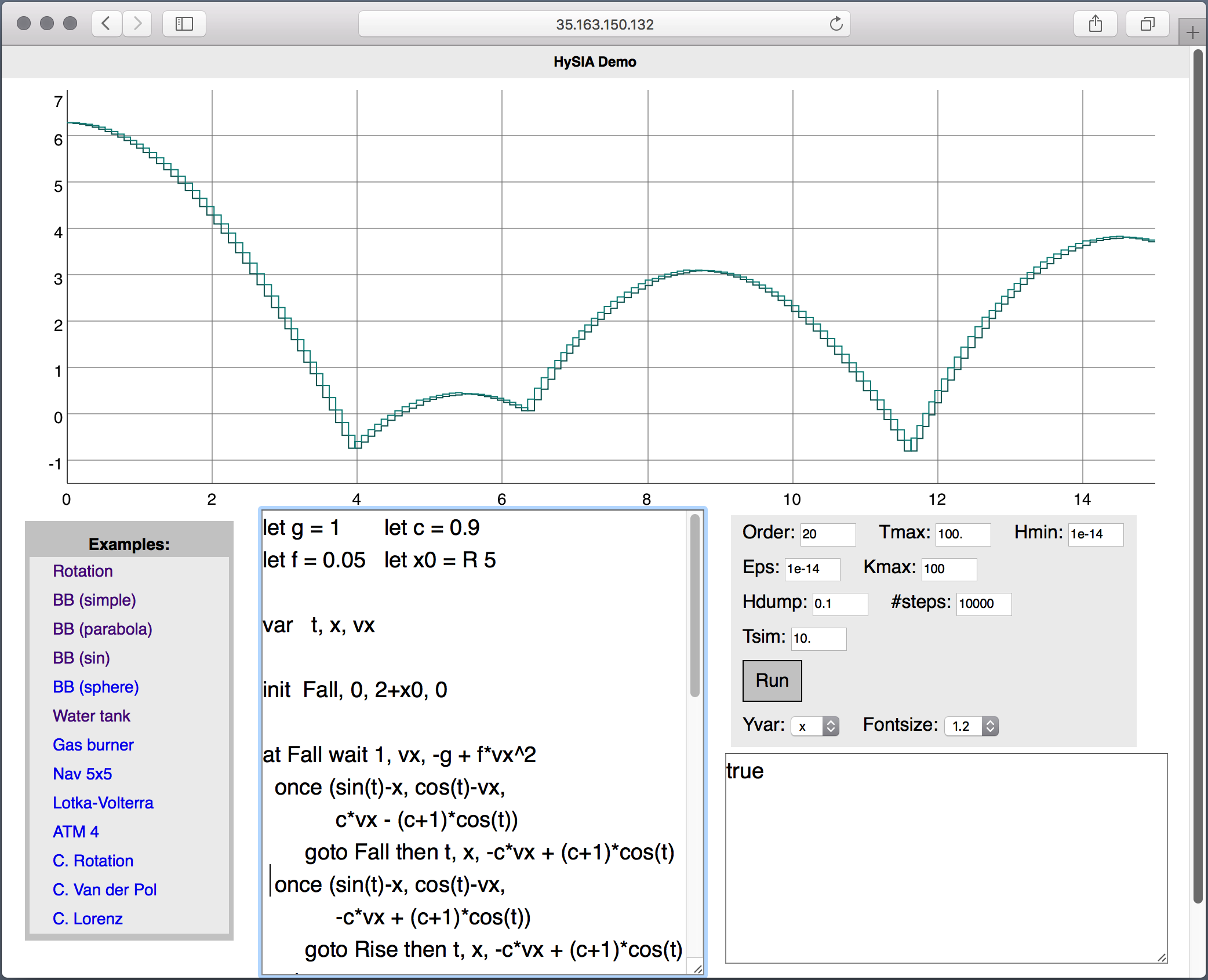}
    \caption{Web interface available at \texttt{https://dsksh.github.io/hysia/}.}
  \label{f:gui}
\end{figure}

The front end contains a parser and a data structure to process hybrid system specifications.
Then, a resulting abstract representation is processed by the simulation module.
After a simulation, the resulting trajectory is processed by the monitoring module to check whether the attached property is satisfied.
Basic interval analysis procedures, e.g., ODE solving, guard evaluation, and jump computation, are implemented in the C++ part.
Both OCaml simulation and monitoring modules are built using these procedures.
HySIA also provides a GUI to users via web browsers (Figure~\ref{f:gui}).

Each component is detailed in the following subsections.

\subsection{Specification Language}
\label{s:input}

\begin{figure}[t]
\lstset{frame=single}
  \begin{lstlisting}
let g = 1    let c = 0.9    let f = 0.05    let x0 = R 5

var    t, x, vx

init   Fall, 0, 2+x0, 0

at Fall wait 1, vx, -g + f*vx^2
    once (sin(t)-x, cos(t)-vx,   c*vx - (c+1)*cos(t))
                       goto Fall then t, x, -c*vx + (c+1)*cos(t)
    once (sin(t)-x, cos(t)-vx, -c*vx + (c+1)*cos(t))
                       goto Rise then t, x, -c*vx + (c+1)*cos(t)
end
at Rise wait 1, vx, -g - f*vx^2
    once (vx, true)  goto Fall then t, x, vx
    once (sin(t)-x, cos(t)-vx)
                       goto Rise then t, x, -c*vx + (c+1)*cos(t)
end

prop   G[0,10] F[0,5] (x-2)
  \end{lstlisting}
  \vspace{-1em}
  \caption{Example of an HA and an STL property.}
  \label{f:input}
\end{figure}

Figure~\ref{f:input} shows an example HA (referred to as \textit{bb-sin} in the sequel) and an STL property described in HySIA's specification language.
This HA models a 1D ball that bounces off a table that is moving sinusoidally; a trajectory of the ball is illustrated in Figures~\ref{f:gui} and \ref{f:traj:sin}.
Line~1 includes the definition of the constants $\mathtt{g}, \mathtt{c}, \mathtt{f}$, and $\mathtt{x0}$; ``\verb|R 5|'' represents a value randomly taken from $[0,5]$ at the execution time.
Line~3 presents the state variables $\mathtt{t}, \mathtt{x}$, and $\mathtt{vx}$.
Line~5 includes the description of the initial location and the initial value for each of the state variables, e.g., $\mathtt{x} := 2 + \mathtt{x0}$.
Lines 7--12 and 13--17 specify the locations \verb|Fall| and \verb|Rise|, respectively.
After the keyword \verb|wait|, the derivative of each state variable is specified.
Moreover, line~7 is interpreted as follows:
\[
    \tfrac{d}{dt}\mathtt{t}(t) = 1, \tfrac{d}{dt}\mathtt{x}(t) = \mathtt{vx}(t), \tfrac{d}{dt}\mathtt{vx}(t) = -\mathtt{g}+\mathtt{f}\ \mathtt{vx}(t)^2.
\]
A sentence starting with \verb|once| describes a location transition; each of the following expressions is the left-hand side of a guard equation and inequalities; line~8 specifies 
\[
    \sin \mathtt{t} - \mathtt{x} = 0 \LAnd \cos \mathtt{t} - \mathtt{vx} > 0 \LAnd
    \mathtt{c}\ \mathtt{vx} - (\mathtt{c}+1) \cos \mathtt{t} > 0.
\]
The expression after \verb|then| specifies the reset of the state
\[
    (\mathtt{t},\mathtt{x},\mathtt{vx}) ~:=~
    (\mathtt{t},\mathtt{x},-\mathtt{c}\ \mathtt{vx} + (\mathtt{c}+1) \cos \mathtt{t}).
\]

The STL property $\Always_{[0,10]}\ \Eventually_{[0,5]}\ \mathtt{x}-2 > 0$ is given in line~19 (see Section~\ref{s:monitoring}).

\subsection{Interval Analysis Procedures}

Every numerical computation in HySIA is performed as a validated interval computation.
Instead of a real value $r \in \RealSet$, we handle a closed interval $[l,u]$, where $l,u \in \FPSet$ ($\FPSet$ denotes the set of floating-point numbers), that encloses $r$ (i.e., $l \leq r \leq u$). Instead of a real vector, we handle a \emph{box} $([l_1,u_1],\ldots, [l_n,u_n])$, i.e., an interval vector, or a \emph{parallelotope}, i.e., a linear transformed box.
A parallelotope is represented as $\langle A, \u, \tilde{x} \rangle$ and interpreted as a region $\{\tilde{x}+Au ~|~ u \in \u \}$, where $A$ is a matrix $\in \FPSet^{n \times n}$, $\u$ is a box, and $\tilde{x}$ is a vector $\in \FPSet^n$.
The CAPD library\footnote{\url{http://capd.ii.uj.edu.pl/}} and the underlying Filib++ library\footnote{\url{http://www2.math.uni-wuppertal.de/~xsc/software/filib.html}} are used for the ODE integration and interval arithmetic, respectively.

\subsection{Simulation Module}

The simulation module iteratively computes a set of parallelotopes, each enclosing a state within the trajectory of the input HA;
it (i) searches for a state that evolves from the initial state and satisfies the guard of a transition, and (ii) computes the next initial state after a discrete transition.
HySIA implements an algorithm that takes into account the wrapping effect that occurs both when integrating an ODE and when computing a discrete transition~\cite{GI2016}.
More precisely, the algorithm is designed based on a consideration that an evolution of an HA state $x$ for the duration $t$, \emph{over a discrete jump}, can be represented as a composite function
\[
    \omega(x,t) ~:=~ \phi_2(\delta(\phi_1(x,\tau(x))), t-\tau(x)),
\]
where $\phi_1$ and $\phi_2$ are continuous trajectories in locations before and after the jump, respectively; $\delta$ is a jump function; $\tau$ is a function that returns the time at which the jump occurs.
The algorithm provides a \emph{parallelotope extension} $\langle \omega \rangle$ of the function $\omega$, i.e., for every simulation time $t$, $\forall x \!\in\! \x,~ \omega(x,t) \in \langle\omega\rangle(\x,t)$ holds, given a parallelotope $\x$.
By iterating this algorithm, we can simulate $k$ jumps from an initial parallelotope $\x_0$ as
\[
    \x_1 := \langle\omega_1\rangle(\x_0,\overline{\tau_1(\x_0)}), ~\ldots, ~
    \x_k := \langle\omega_k\rangle(\x_{k-1},\overline{\tau_k(\x_{k-1})}),
\]
where $\overline{\tau_i(\x_{i-1})}$ represents the upper bound of the time interval.

Involving interval values within a parallelotope extension is straightforward;
therefore, HySIA allows input HA to involve an interval value within the specification.
For instance, the initial value of a trajectory of an HA can be parameterized by an interval vector.
Using the interval analysis techniques, a value of the parallelotope extension is verified to contain a unique state of the trajectory for each initial value.
Overall, the verification process of HySIA proves
\[
    \ForAll{x_0}{\x_0}, ~
    \ForAll{i}{\{1,\ldots,k\}}, ~
    \ExistsU{x_i}{\langle\omega_i\rangle(\x_{i-1},\overline{\tau_i(\x_{i-1})})}, ~ x_i = \omega_i(x_{i-1},\overline{\tau_i(\x_{i-1})}).
\]

\subsection{Monitoring Module}

The monitoring module evaluates an STL formula based on both boolean-valued~\cite{Maler2003} and real-valued (i.e., robustness)~\cite{Donze2010} semantics.
The procedure implemented in HySIA is incomplete: it outputs either $\Valid$, $\Unsat$, or \emph{unknown}.
Given an HA and an STL property, the output $\Valid$ implies that every trajectory of the HA satisties the property;
the output $\Unsat$ implies that no trajectory of the HA satisfies the property.
The inconclusive result is obtained either when
(i) the simulation module fails to verify the unique existence of a solution trajectory or
(ii) the result is affected by a precise boundary interval as described below.

For boolean-valued monitoring, (interval enclosures of) zero-crossing points for each atomic proposition $p$ in the STL formula are tracked by the simulation module.
As a result, a list of time intervals is obtained, within which $p$ holds.
The bounds of the time intervals are represented as intervals, where the satisfiability of $p$ is unknown.
Then, the monitoring module processes the lists according to the STL formula in a bottom-up fashion to check whether the STL formula holds at time 0~\cite{Maler2003,Ishii2016IEICE}; 
the module outputs unknown when a boundary interval intersects time 0.

For real-valued semantics, the monitoring module performs the dedicated algorithm~\cite{Donze2013}, which is extended to handle the interval overapproximation of signals.
The current version of the module supports the untimed portion of STL evaluated with bounded-length trajectories.
In the monitoring process, the module maintains a list of objects, each of which represents an enclosure of a segment of the resulting signal, where the signal value changes monotonically. Again the list may contain small segments where the monotonicity is unknown because of the overapproximation.

\section{Examples}

In this section, we show the basic functionalities of HySIA: simulation and monitoring of HA.
The reported experiments were run using a 2.7GHz Intel Core i5 processor with 16GB RAM.
See the documentation\footnote{\url{https://dsksh.github.io/hysia/manual.pdf}} for more details.

\subsection{Validated Numerical Simulation}

\subsubsection{Simulation of nonlinear HA.}

\begin{figure}[t]
\centering
  \begin{minipage}{.63\textwidth}
    \centering
    \includegraphics[width=\linewidth]{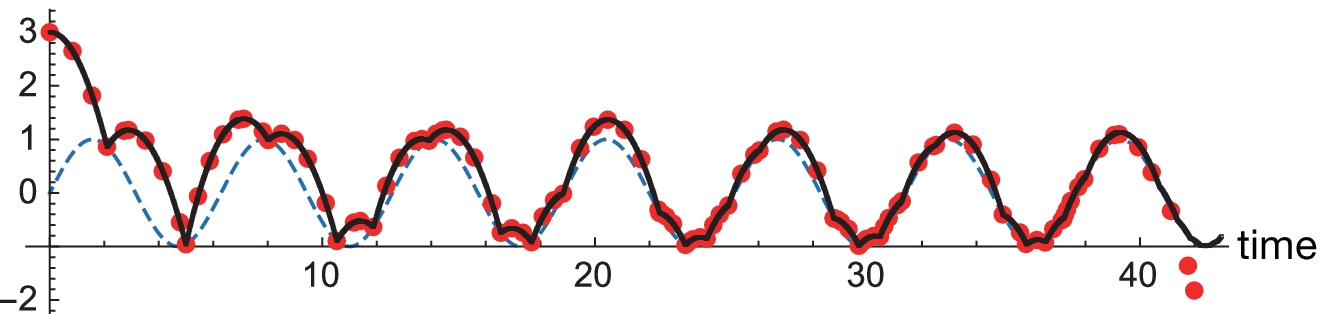}

    \vspace*{.5em}

    \includegraphics[width=.98\linewidth]{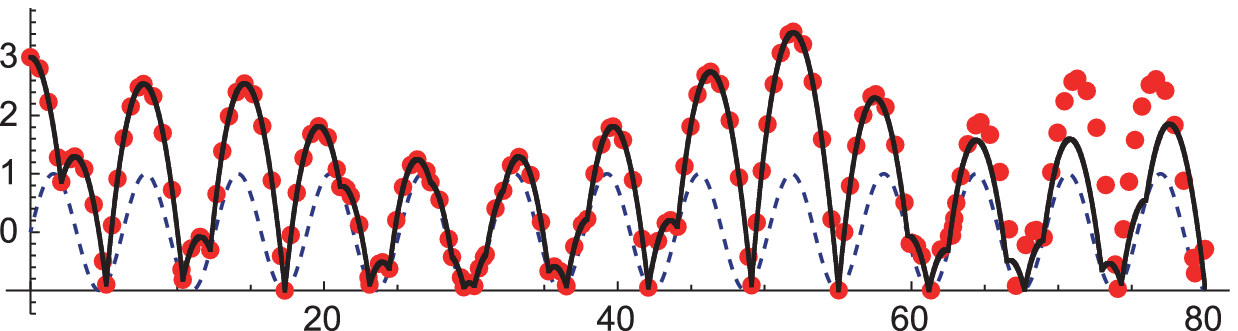}

    \caption{Trajectories for the variable $\mathtt{x}$ of two \textit{bb-sin} instances with the initial value $\mathtt{x}=3$; the constant $\mathtt{c}$ is set as 1 in the experiment below. Solid lines are enclosures computed by HySIA, dots are values computed by MATLAB/Simulink, and dashed lines are the solutions of the guard equations.}
    \label{f:traj:sin}
  \end{minipage}
  \hspace{1em}
  \begin{minipage}{.33\textwidth}
    \includegraphics[width=\linewidth]{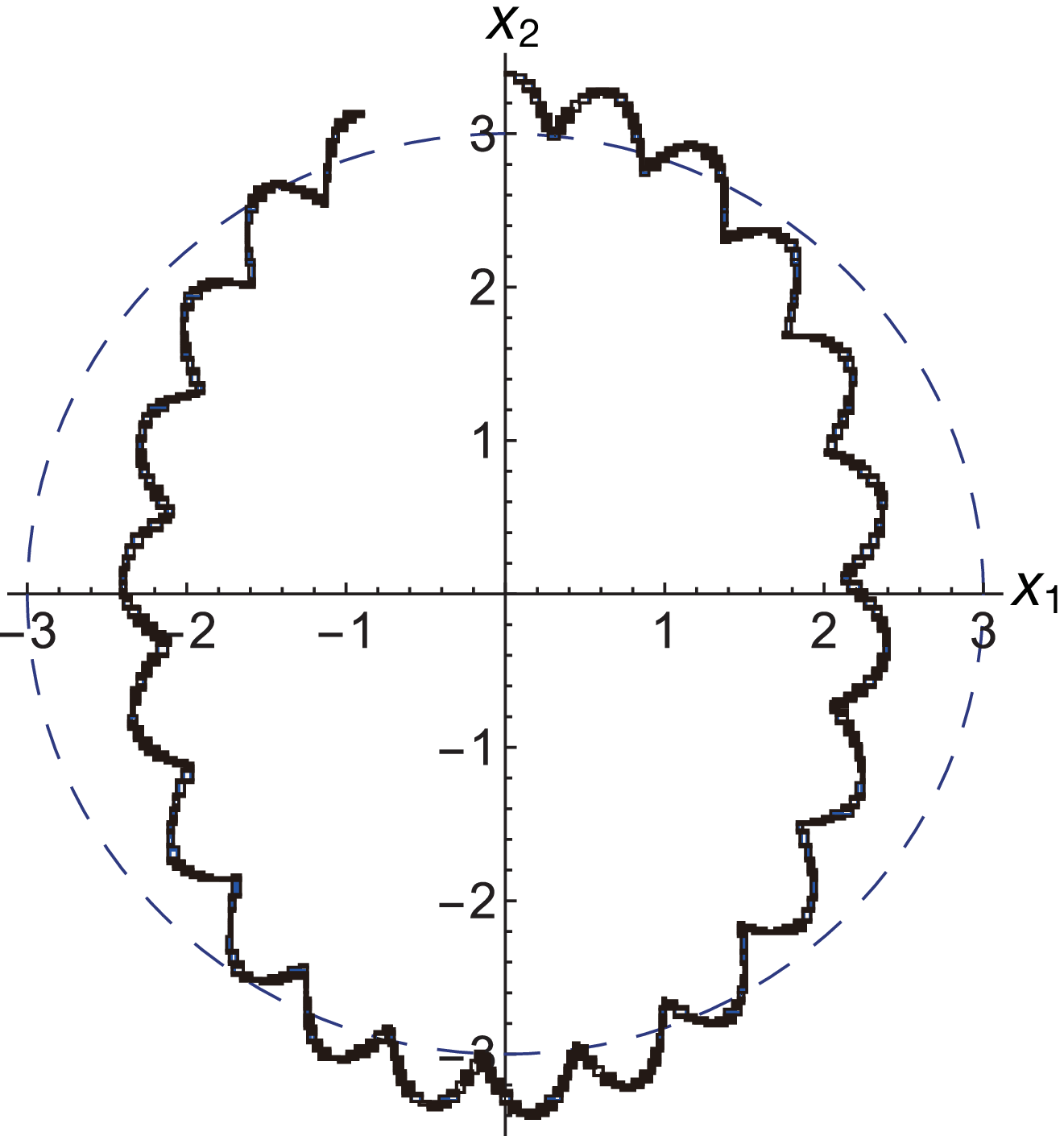}
  \caption{Trajectory of $\textit{bb-sph}_{3.4}$,
      where the initial value is set as $(0,0,3.4,1,1,0)$. Thick and dashed lines represent the computed enclosures and the outline of the sphere, respectively.}
  \label{f:traj:sphere}
  \end{minipage}
\end{figure}

HySIA correctly simulates nonlinear HA such as the \textit{bb-sin} example in Section~\ref{s:input}.
Figure~\ref{f:traj:sin} shows simulation results of the two instances of \textit{bb-sin} (Figure~\ref{f:input}) computed with HySIA and MATLAB/Simulink (R2016b; we use the simulator ``ODE45'' and refine the minimum step size to $10^{-8}$). The two results for each instance differ as simulation time proceeds indicating that nonvalidated numerical methods may output a wrong result.

Another example is the \textit{lotka-volterra} system~\cite{GI2016} that switches two nonlinear ODEs.
We can also consider a 3D \textit{bb-sph} example~\cite{GI2016}, in which a ball bounces off a sphere based on the universal gravitation between the ball and the sphere.
Its ODE and the guard can be modeled as
\[
    \frac{d^2}{dt^2}x(t) = -\frac{x(t)}{|\!| x(t) |\!|_2^3}, \qquad
    g(x(t)) \ \equiv\ |\!| x(t) |\!|_2^2 - r^2 = 0 \LAnd \frac{d}{dt}x(t) \cdot \nabla (|\!| x(t) |\!|_2^2) > 0,
\]
where $x$ represents the position of the ball and $\nabla$ is the gradient operator.
A simulation result for this example is shown in Figure~\ref{f:traj:sphere}.

\begin{table}[t]
\centering
  \caption{\label{t:sim} Simulation results.
Each column shows the problem, the dimension of the state variable, the number of jumps HySIA can simulate, the number of jumps a naive box-enclosure method can simulate, and the average CPU time taken for a continuous phase and a jump.}
  \begin{tabular}{l|c|r|r|r} \hline \hline
    problem & dim. & \# jumps & box & time \\
    \hline
    $\textit{bb-sin}$	        &  2 &  1433 & 24 & $<0.1$s \\
    $\textit{navigation}$	    &  4 & 24731 & 51 & $<0.1$s \\
    $\textit{lotka-volterra}$	&  2 &  1875 & 59 & $<0.1$s \\
    $\textit{bb-sph}_{3.4}$	&  6 &   285 & 14 & $0.29$s \\
    $\textit{bb-sph}_{3.1}$	&  6 &   554 & 17 & $0.14$s \\
    $\textit{bb-sph}_{3.1+[-10^{-8},10^{-8}]}$ &  6 & 86 & 9 & $0.14$s \\
    \hline
  \end{tabular}
\end{table}

\subsubsection{Simulation for a long duration.}

HySIA is able to simulate a number of jumps for input HA with its reduction mechanism of the wrapping effect.
If the possible number of jumps is unbounded, the enclosure of the trajectory enlarges as the simulation proceeds, and the simulation results in an error due to failure of the verification process.
Table~\ref{t:sim} illustrates the results of several simulation experiments.
In the experiments, HySIA simulates more jumps compared to the naive interval-based method; it outperforms other reachability tools when the interval enclosure of the state is sufficiently tight.
%
The computation time is efficient and most of the timing is taken by the ODE integration process; a \emph{stiff} ODE of larger dimension may require a longer computation time.

\subsubsection{Simulation of HA involving interval values.}

HySIA allows an HA to involve an interval value within the model;
in the last experiment reported in Table~\ref{t:sim}, an interval value is set as the initial height.
However, putting an interval in the model easily make the verification process difficult;
as a result, the number of simulated jumps decreases in the experiment.

\subsection{STL Property Monitoring}
\label{s:monitoring}

\begin{table}[t]
  \centering
    \caption{\label{t:mon} Monitoring results of the STL formula $\Always_{[0,t]} \Eventually_{[0,5]} \mathtt{x}\!-\!2 \!>\! 0$ against the \textit{bb-sin} example.
Each column shows the value $t$, the width of an interval initial value of $\mathtt{x}$, the numbers of runs for each output value, and the average CPU time taken for a run. 
    } 
  \begin{tabular}{l|r|r|r|r|r} \hline \hline
    $t$ 
    & width & \# $\mathrm{valid}$ & \# $\mathrm{unsat}$ & \# $\mathrm{unknown}$ & time \\
	\hline
    10
	& 0\ & 238 & 762 & 0 & 0.2s \\
    10
	& 0.01\ & 123 & 10 & 867 & 0.2s \\
    100 
	& 0\ & 134 & 17 & 849 & 0.9s \\
	\hline
  \end{tabular}
\end{table}

Table~\ref{t:mon} shows experimental results 
of boolean-valued monitoring of
the STL property against the \textit{bb-sin} example.
The property
\[
    \Always_{[0,t]}\ \Eventually_{[0,5]}\ \mathtt{x}\!-\!2 \!>\! 0 ~~\equiv~~
      \neg ( \True\ \Until_{[0,t]}\ \True\ \Until_{[0,5]}\ \mathtt{x} - 2 > 0),
\]
where $t$ is set as either 10 or 100, means that the value of $\mathtt{x}$ exceeds 2 within every 5 time units for the initial $t$ time units.
In the three experiments, 1000 runs are performed with different settings.
HySIA evaluates the property to $\Valid$ or $\Unsat$ only when the result is reliable.
In the first experiment, the monitoring process successfully checks whether or not the property holds; we count the number of outputs $\Valid$ and $\Unsat$ in Table~\ref{t:mon}
(the result differs from that in \cite{Ishii2015-NSV} because the value of \verb|f| is different and the verification process is slightly modified).
When a monitoring run is badly conditioned, so that the verification process in the monitoring process fails, HySIA will output $\Unknown$ (or terminate with an error information for some cases).
In the second experiment, when an initial value of $\mathtt{x}$ is set as an interval of $0.01$ width, we obtain the result $\Unknown$ for 867 times.
In the third experiment, we set $t$ as 100 to monitor for 105 time units; then we have the result $\Unknown$ for 849 times.
Even though the $\Unknown$ results are inconclusive, we consider this verification mechanism is valuable for monitoring a system reliably and efficiently (cf. timings in the last column).

\begin{figure}[t]
\centering
\includegraphics[width=.8\linewidth]{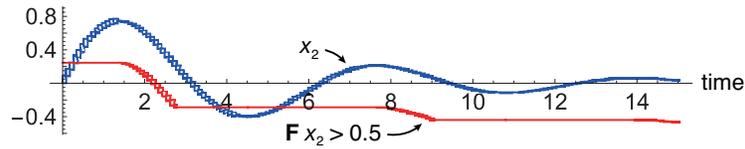}
\caption{Interval enclosure of a robustness signal of $\Eventually x_2 > 0.5$.}
\label{f:robustness}
\end{figure}

HySIA provides a real-valued monitoring feature for continuous systems, i.e., HA consisting of a location and no transitions.
HySIA can compute a robustness signal for unbounded STL properties within a given bounded time horizon.
Figure~\ref{f:robustness} illustrates a computed signal of a property of a simple rotation system.

\section{Conclusion and Future Work}

The HySIA tool is presented.
A web demonstation is available at \url{https://dsksh.github.io/hysia/}.
We consider that HySIA is a promising testbed for reliable runtime verification of nonlinear hybrid systems.
The whole process of HySIA for simulation and monitoring is implemented using various interval analysis techniques.
The tool is able to simulate and monitor various HA and properties for reasonable duration, with a computational efficiency.

As a future work, more detailed analysis and explanation of the inconclusive results will be needed.
Extension of HySIA can be planned to incorporate reliable simulation into various runtime verifications such as statistical model checking and conformance testing of hybrid systems.

\paragraph{Acknowledgments.}
This work was partially funded by JSPS (KAKENHI 25880008, 15K15968, and 26280024).

\bibliographystyle{splncs03}
\bibliography{article}


\end{document}